\documentclass[10pt]{article}
\usepackage{graphicx}
\usepackage{amsmath}
\usepackage{amssymb}
\usepackage{caption2}
\usepackage[figuresright]{rotating}
\begin{document}
\begin{center}
{\large {\bf \sc{  Analysis of the vector hidden-charm tetraquark  states without explicit P-waves via the QCD sum rules }}} \\[2mm]
Zhi-Gang  Wang \footnote{E-mail: zgwang@aliyun.com.  }     \\
 Department of Physics, North China Electric Power University, Baoding 071003, P. R. China
\end{center}

\begin{abstract}
In the present work,  we adopt the   scalar, pseudoscalar, vector,  axialvector and tensor  (anti)diquark operators  as the elementary  building blocks     to construct  vector and tensor local  four-quark currents without introducing explicit P-waves,  and explore the mass spectrum of the vector hidden-charm tetraquark states via the QCD sum rules comprehensively,  and revisit the interpretations of the existing  $Y$ states in the  scenario of vector tetraquark  states. We resort to  the energy scale formula to enhance the pole contributions and improve the convergent behaviors of the operator product expansion, and we should bear in mind that the predictions are rather sensitive to the particular  energy scales which obey the uniform/same constraint.
The  predicted vector hidden-charm tetraquark states can be confronted to the experimental data  in the future.
 \end{abstract}



\section{Introduction}
In recent years, several vector  charmonium-like states have been observed, such as the $Y(4008)$, $Y(4160)$,
$Y(4220)$, $Y(4230)$, $Y(4260)$, $Y(4320)$, $Y(4360)$, $Y(4390)$, $Y(4630)$, $Y(4660)$, etc,  they  cannot be accommodated suitably   in the conventional two quark model,  we have to introduce additional quark or gluon degrees of freedom to explore their properties \cite{XYZ-review}.
For example, the $Y(4260)$ observed by the BaBar collaboration \cite{BaBar4260-0506}, the $Y(4220)$, $Y(4390)$ and $Y(4320)$ observed  by the BESIII collaboration  \cite{BES-Y4390,BES-Y4220-Y4320}, and the $Y(4360)$, $Y(4660)$, $Y(4630)$ observed by the
  Belle collaboration \cite{Belle4660-0707-1,Belle4660-0707-2,Belle4630-0807} are excellent candidates for the vector tetraquark states.  Considering  the analogous masses and decay widths,  we can take the $Y(4260)$, $Y(4220)$ and $Y(4230)$ as the same meson, the $Y(4360)$ and $Y(4320)$ as the same meson, and the $Y(4660)$ and $Y(4630)$ as the same meson \cite{XYZ-review}.

The QCD sum rules method is  a vigorous  theoretical tool in exploring the properties of the exotic $X$, $Y$ and $Z$ states, there have been several    possible interpretations of the existing $Y$ states according to the investigations via the QCD sum rules, such as the vector tetraquark states \cite{Nielsen-4260-4460,ChenZhu-Vector-Axial,WangY4360Y4660-1803,Wang-tetra-formula,WangEPJC-1601-Mc,ZhangHuang-PRD,
ZhangHuang-JHEP,Vector-Tetra-WZG-P-wave-1,Vector-Tetra-WZG-P-wave,Vector-Tetra-WZG-4100,Vector-4660-Azizi}, the tetraquark molecular states (color-singlet-color-singlet type tetraquark states) \cite{Vector-mole-WZG-CPC}, the charmonium-tetraquark mixing states \cite{Vector-mix-Nielsen-4260}.

In the scenario of tetraquark states, we can construct the color-antitriplet-color-triplet type, color-sextet-color-antisextet  type, color-singlet-color-singlet type and color-octet-color-octet  type  four-quark configurations to explore the tetraquark properties. All those four-quark configurations are compact objects as we choose the local currents to interpolate them in the QCD sum rules.  While  we usually prefer the diquarks  in the color-antitriplet to other configurations in selecting  the basic building blocks, because the attractive interactions induced by one-gluon exchange favor forming diquarks in the color-antitriplet \cite{One-gluon-1,One-gluon-2}.

The diquark operators $\varepsilon^{ijk}q^{T}_j C\Gamma q^{\prime}_k$ in the color-antitriplet have  five  structures  in Dirac spinor space, where the $i$, $j$ and $k$ are color indexes,  $C\Gamma=C\gamma_5$, $C$, $C\gamma_\mu \gamma_5$,  $C\gamma_\mu $ and $C\sigma_{\mu\nu}$ for the scalar, pseudoscalar, vector, axialvector  and  tensor diquarks, respectively.
In the  non-relativistic quark model, an additional P-wave can change the parity by contributing a factor $(-)^L=-$,   where  the angular momentum $L=1$.
The $C\gamma_5$ and $C\gamma_\mu$ diquark states have the spin-parity $J^P=0^+$ and $1^+$, respectively, the corresponding  $C$ and $C\gamma_\mu\gamma_5$ diquark states have the spin-parity $J^P=0^-$ and $1^-$, respectively, the net effects of the  P-waves   are embodied in the underlined  $\gamma_5$ in the  $C\gamma_5 \underline{\gamma_5} $ and $C\gamma_\mu \underline{\gamma_5} $.  The tensor diquark states have both the $J^P=1^+$ and $1^-$ components, we project out the $1^+$ and $1^-$ components explicitly in one way or the other, and introduce the symbols $\widetilde{A}$ and $\widetilde{V}$ to represent the corresponding axialvector and vector diquark operators, respectively.
We can also introduce the  P-wave explicitly in the  $C\gamma_5$ and $C\gamma_\mu$ diquark operators and acquire the vector diquark operators  $\varepsilon^{ijk}q^{T}_j C\gamma_5 \stackrel{\leftrightarrow}{\partial}_\mu q^{\prime}_k$ or tensor diquark operators  $\varepsilon^{ijk}q^{T}_j C\gamma_\mu \stackrel{\leftrightarrow}{\partial}_\nu q^{\prime}_k$, where the derivative  $\stackrel{\leftrightarrow}{\partial}_\mu=\stackrel{\rightarrow}{\partial}_\mu-\stackrel{\leftarrow}{\partial}_\mu$ embodies  the P-wave effects.

To explore the ground state  hadron masses and make possible interpretations of the $Y$ states, we can adopt  an S-wave diquark (antidiquark) and a P-wave antidiquark (diquark) pair as the basic building blocks  to construct the vector hidden-charm tetraquark states without introducing an explicit P-wave, or introduce an explicit P-wave between the S-wave diquark and  antidiquark constituents to construct the vector hidden-charm tetraquark states, and investigate them with the QCD sum rules in a comprehensive way to avoid possible  biased  analyses.

The predictions depend heavily on the input parameters chosen  at the QCD side of the QCD sum rules, even in the same scenario of the tetraquark states, the same interpolating currents lead to quite different interpretations of the $Y$ states  \cite{Nielsen-4260-4460,ChenZhu-Vector-Axial,WangY4360Y4660-1803}. A comprehensive analysis with the same input parameters and same (or uniform) treatments is necessary, although all the $X$, $Y$ and $Z$ states have been investigated  with the QCD sum rules in one way or the other.

In Refs.\cite{Vector-Tetra-WZG-P-wave-1,Vector-Tetra-WZG-P-wave}, we introduce an explicit  P-wave between the diquark and antidiquark constituents to construct the  vector and tensor local four-quark currents,  and explore the  vector tetraquark states via the QCD sum rules systematically  by calculating  the vacuum condensates up to dimension 10 in a consistent way, and prefer  the modified  energy scale formula $\mu=\sqrt{M^2_{X/Y/Z}-(2{\mathbb{M}}_c+0.5\,\rm{GeV})^2}$ with the effective $c$-quark mass ${\mathbb{M}}_c$  to select  the pertinent  energy scales of the QCD spectral densities considering the explicit P-wave, and acquire the lowest masses of the vector tetraquark states up to now and reexamine the possible interpretations of all the existing  $Y$ states.  The  predictions support  identifying the
 $Y(4220/4260)$,  $Y(4320/4360)$, $Y(4390)$ and $Z(4250)$ as the vector tetraquark   states with a relative P-wave between the diquark and antidiquark pair, see Table \ref{Assignment-Vector-P}. In Table \ref{Assignment-Vector-P}, we show the quantum numbers of the $Y$ states
in the non-relativistic  diquark-antidiquark model explicitly, where the $L$ is the angular momentum between the diquark and antidiquark constituents, the total spin $\vec{S}=\vec{S}_{qc}+ \vec{S}_{\bar{q}\bar{c}}$, and the total angular momentum $\vec{J}=\vec{S}+ \vec{L}$.

In the phenomenological  type-II diquark-antidiquark model \cite{Maiani-II-type},  L. Maiani et al identify  the $Y(4008)$, $Y(4260)$, $Y(4290/4220)$  and $Y(4630)$ as  the four ground  states with  $L=1$ based on the effective  Hamiltonian with the spin-spin and spin-orbit  interactions but neglect the spin-spin interactions between the quarks and antiquarks. In Ref.\cite{Ali-Maiani-Y}, A. Ali et al incorporate the dominant spin-spin, spin-orbit and tensor interactions, and observe that the preferred  interpretations of the ground states with  $L=1$ are the $Y(4220)$, $Y(4330)$, $Y(4390)$, $Y(4660)$, which are also shown explicitly in Table \ref{Assignment-Vector-P}.

 From Table \ref{Assignment-Vector-P}, we can see clearly that the $Y(4660)$ cannot be identified as  the diquark-antidiquark type vector tetraquark state with a relative P-wave between the diquark and antidiquark constituents according to the calculations based on the QCD sum rules \cite{Vector-Tetra-WZG-P-wave}, which differs  from Ref.\cite{Ali-Maiani-Y} remarkably. More works on the $Y(4660)$ based on the QCD sum rules are still needed.

\begin{table}
\begin{center}
\begin{tabular}{|c|c|c|c|c|c|c|c|}\hline\hline
$|S_{qc}, S_{\bar{q}\bar{c}}; S, L; J\rangle$                                &$M_Y(\rm{GeV})$     &\cite{Vector-Tetra-WZG-P-wave}  &\cite{Ali-Maiani-Y} \\ \hline
$|0, 0; 0, 1; 1\rangle$                                                      &$4.24\pm0.10$       &$Y(4220)$                       &$Y(4220)$            \\ \hline

$\frac{1}{\sqrt{2}}\left(|1, 0; 1, 1; 1\rangle+|0, 1; 1, 1; 1\rangle\right)$ &$4.31\pm0.10$       &$Y(4320/4390)$                  &$Y(4330)$             \\ \hline

$|1, 1; 0, 1; 1\rangle$                                                      &$4.28\pm0.10$       &$Y(4220/4320)$                  &$Y(4390)$            \\ \hline
$|1, 1; 2, 1; 1\rangle$                                                      &$4.33\pm0.10$       &$Y(4320/4390)$                  &$Y(4660)$         \\ \hline \hline
\end{tabular}
\end{center}
\caption{ The masses  of the vector tetraquark states from the QCD sum rules \cite{Vector-Tetra-WZG-P-wave}, and possible interpretations of the $Y$ states based on the QCD sum rules \cite{Vector-Tetra-WZG-P-wave} and diquark-antidiquark model \cite{Ali-Maiani-Y}.   }\label{Assignment-Vector-P}
\end{table}

Now let us go back to the vector tetraquark states without introducing the explicit P-waves. In Ref.\cite{WangY4360Y4660-1803}, we construct the $C \otimes \gamma_\mu C$ and $C\gamma_5 \otimes \gamma_5\gamma_\mu C$ type  four-quark currents with the $J^{PC}=1^{--}$ to interpolate  the vector  tetraquark states,  and acquire  four QCD sum rules. We use the energy scale  formula $\mu=\sqrt{M^2_{Y}-(2{\mathbb{M}}_c)^2}$ to select   the pertinent energy scales of the QCD spectral densities considering the P-wave is embodied implicitly in the (anti)diquarks, and adopt the experimental masses of the $Y(4260/4220)$, $Y(4360/4320)$, $Y(4390)$ and $Y(4660/4630)$ as input parameters and  make great efforts to fit the pole residues to reproduce the QCD sum rules at the quark level. The numerical results support identifying  the $Y(4660/4630)$  as the  $C \otimes \gamma_\mu C$ type vector tetraquark state $c\bar{c}s\bar{s}$, identifying the $Y(4360/4320)$ as $C\gamma_5 \otimes \gamma_5\gamma_\mu C$  type vector tetraquark state $c\bar{c}q\bar{q}$,  and disfavor identifying the $Y(4260/4220)$ and $Y(4390)$ as the pure vector tetraquark states without introducing mixing effects or  explicit relative P-waves.
Our calculations based on the QCD sum rules are consistent with each other \cite{WangY4360Y4660-1803,Vector-Tetra-WZG-P-wave}.

In fact, we can also construct other local four-quark currents to interpolate the vector hidden-charm tetraquark states, for example, the $S\tilde{V}-\tilde{V}S$ type hidden-charm tetraquark state with the $J^{PC}=1^{--}$ \cite{Vector-Tetra-WZG-4100}.  It is interesting and valuable to exhaust all the vector four-quark configurations, and investigate the ground state mass spectrum of the vector tetraquark states with the  pseudoscalar ($P$), scalar ($S$), axialvector ($A$, $\tilde{A}$) and vector ($V$, $\tilde{V}$) (anti)diquark operators without introducing the relative P-wave explicitly.

In Ref.\cite{WZG-HC-spectrum-PRD}, we adopt the scalar, pseudoscalar,  axialvector, vector, tensor (anti)diquark operators as the basic constituents, and construct the scalar, axialvector and tensor four-quark currents to explore  the  mass spectrum of the ground state hidden-charm tetraquark states  via
the QCD sum rules  comprehensively, and revisit the interpretations of the $X$, $Y$, $Z$ states in the  scenario of tetraquark  states in a consistent way.  In Ref.\cite{WZG-Zcs-spectrum-CPC}, we adopt  the scalar and axialvector (anti)diquark operators  as the elementary   constituents  to construct the four-quark  currents and   investigate   the  axialvector tetraquark states $c\bar{c}u\bar{s}$ and  take account of the light flavor   $SU(3)$ mass-breaking effect to estimate the mass spectrum of the  hidden-charm  tetraquark states having  the strangeness considering   our previous  works \cite{WZG-HC-spectrum-PRD}. In all the works \cite{WZG-HC-spectrum-PRD,WZG-Zcs-spectrum-CPC}, we resort to the energy scale formula  $\mu=\sqrt{M^2_{X/Y/Z}-(2{\mathbb{M}}_c)^2}$ to select the pertinent  energy scales of the QCD spectral densities to enhance  the pole contributions and improve the convergent behaviors of the operator product expansion \cite{Wang-tetra-formula}.

In  the present work,  we adopt the scalar, pseudoscalar, vector, axialvector and tensor  (anti)diquark operators  as the elementary constituents  to construct  four-quark vector and tensor currents without introducing the explicit P-wave (in other words, the P-wave is embodied implicitly in the negative-parity of the diquarks),  and investigate  the mass spectrum of the vector hidden-charm tetraquark states via the QCD sum rules comprehensively by using  the energy scale formula  $\mu=\sqrt{M^2_{X/Y/Z}-(2{\mathbb{M}}_c)^2}$ to determine  the pertinent  energy scales of the QCD spectral densities,  and revisit the interpretations of the existing $Y$ states in the  scenario of vector tetraquark  states  and  make great efforts to accommodate the exotic $Y$ states as many as possible in a consistent way.

The article is arranged as follows:  we derive the QCD sum rules for the masses and pole residues  of  the vector hidden-charm tetraquark states in section 2; in section 3, we   present the numerical results and discussions; section 4 is reserved for our conclusion.

\section{QCD sum rules for  the  vector hidden-charm  tetraquark states}
Firstly let us  write down  the two-point correlation functions  $\Pi_{\mu\nu}(p)$ and $\Pi_{\mu\nu\alpha\beta}(p)$,
\begin{eqnarray}\label{CF-Pi}
\Pi_{\mu\nu}(p)&=&i\int d^4x e^{ip \cdot x} \langle0|T\Big\{J_\mu(x)J_{\nu}^{\dagger}(0)\Big\}|0\rangle \, ,\nonumber\\
\Pi_{\mu\nu\alpha\beta}(p)&=&i\int d^4x e^{ip \cdot x} \langle0|T\Big\{J_{\mu\nu}(x)J_{\alpha\beta}^{\dagger}(0)\Big\}|0\rangle \, ,
\end{eqnarray}
where the currents
\begin{eqnarray}
J_\mu(x)&=&J^{PA}_{-,\mu}(x)\, ,\,\, J^{PA}_{+,\mu}(x)\, , \,\,J^{SV}_{-,\mu}(x)\, , \,\, J^{SV}_{+,\mu}(x)\, ,\,\,
J_{-,\mu}^{\widetilde{V}A}(x)\, ,\,\, J_{+,\mu}^{\widetilde{V}A}(x)\, , \,\, J_{-,\mu}^{\widetilde{A}V}(x)\, , \,\, J_{+,\mu}^{\widetilde{A}V}(x)\, , \nonumber\\
J_{\mu\nu}(x)&=&J^{S\widetilde{V}}_{-,\mu\nu}(x)\, , \,\, J^{S\widetilde{V}}_{+,\mu\nu}(x)\, , \,\, J^{P\widetilde{A}}_{-,\mu\nu}(x)\, , \,\,
J^{P\widetilde{A}}_{+,\mu\nu}(x)\, , \, \, J^{AA}_{-,\mu\nu}(x)\, ,
\end{eqnarray}
\begin{eqnarray}
J^{PA}_{-,\mu}(x)&=&\frac{\varepsilon^{ijk}\varepsilon^{imn}}{\sqrt{2}}\Big[u^{Tj}(x)Cc^k(x) \bar{d}^m(x)\gamma_\mu C \bar{c}^{Tn}(x)-u^{Tj}(x)C\gamma_\mu c^k(x)\bar{d}^m(x)C \bar{c}^{Tn}(x) \Big] \, ,\nonumber\\
J^{PA}_{+,\mu}(x)&=&\frac{\varepsilon^{ijk}\varepsilon^{imn}}{\sqrt{2}}\Big[u^{Tj}(x)Cc^k(x) \bar{d}^m(x)\gamma_\mu C \bar{c}^{Tn}(x)+u^{Tj}(x)C\gamma_\mu c^k(x)\bar{d}^m(x)C \bar{c}^{Tn}(x) \Big] \, ,\nonumber\\
J^{SV}_{-,\mu}(x)&=&\frac{\varepsilon^{ijk}\varepsilon^{imn}}{\sqrt{2}}\Big[u^{Tj}(x)C\gamma_5c^k(x) \bar{d}^m(x)\gamma_5\gamma_\mu C \bar{c}^{Tn}(x)+u^{Tj}(x)C\gamma_\mu\gamma_5 c^k(x)\bar{d}^m(x)\gamma_5C \bar{c}^{Tn}(x) \Big] \, ,\nonumber\\
J^{SV}_{+,\mu}(x)&=&\frac{\varepsilon^{ijk}\varepsilon^{imn}}{\sqrt{2}}\Big[u^{Tj}(x)C\gamma_5c^k(x) \bar{d}^m(x)\gamma_5\gamma_\mu C \bar{c}^{Tn}(x)-u^{Tj}(x)C\gamma_\mu\gamma_5 c^k(x)\bar{d}^m(x)\gamma_5C \bar{c}^{Tn}(x) \Big] \, ,\nonumber\\
\end{eqnarray}

\begin{eqnarray}
J_{-,\mu}^{\widetilde{V}A}(x)&=&\frac{\varepsilon^{ijk}\varepsilon^{imn}}{\sqrt{2}}\Big[u^{Tj}(x)C\sigma_{\mu\nu} c^k(x)\bar{d}^m(x)\gamma^\nu C \bar{c}^{Tn}(x)-u^{Tj}(x)C\gamma^\nu c^k(x)\bar{d}^m(x)\sigma_{\mu\nu} C \bar{c}^{Tn}(x) \Big] \, , \nonumber\\
J_{+,\mu}^{\widetilde{V}A}(x)&=&\frac{\varepsilon^{ijk}\varepsilon^{imn}}{\sqrt{2}}\Big[u^{Tj}(x)C\sigma_{\mu\nu} c^k(x)\bar{d}^m(x)\gamma^\nu C \bar{c}^{Tn}(x)+u^{Tj}(x)C\gamma^\nu c^k(x)\bar{d}^m(x)\sigma_{\mu\nu} C \bar{c}^{Tn}(x) \Big] \, , \nonumber\\
J_{-,\mu}^{\widetilde{A}V}(x)&=&\frac{\varepsilon^{ijk}\varepsilon^{imn}}{\sqrt{2}}\Big[u^{Tj}(x)C\sigma_{\mu\nu}\gamma_5 c^k(x)\bar{d}^m(x)\gamma_5\gamma^\nu C \bar{c}^{Tn}(x)+u^{Tj}(x)C\gamma^\nu\gamma_5 c^k(x)\bar{d}^m(x)\gamma_5\sigma_{\mu\nu} C \bar{c}^{Tn}(x) \Big] \, , \nonumber\\
J_{+,\mu}^{\widetilde{A}V}(x)&=&\frac{\varepsilon^{ijk}\varepsilon^{imn}}{\sqrt{2}}\Big[u^{Tj}(x)C\sigma_{\mu\nu}\gamma_5 c^k(x)\bar{d}^m(x)\gamma_5\gamma^\nu C \bar{c}^{Tn}(x)-u^{Tj}(x)C\gamma^\nu\gamma_5 c^k(x)\bar{d}^m(x)\gamma_5\sigma_{\mu\nu} C \bar{c}^{Tn}(x) \Big] \, , \nonumber\\
\end{eqnarray}

\begin{eqnarray}
J^{S\widetilde{V}}_{-,\mu\nu}(x)&=&\frac{\varepsilon^{ijk}\varepsilon^{imn}}{\sqrt{2}}\Big[u^{Tj}(x)C\gamma_5 c^k(x)  \bar{d}^m(x)\sigma_{\mu\nu} C \bar{c}^{Tn}(x)- u^{Tj}(x)C\sigma_{\mu\nu} c^k(x)  \bar{d}^m(x)\gamma_5 C \bar{c}^{Tn}(x) \Big] \, , \nonumber\\
J^{S\widetilde{V}}_{+,\mu\nu}(x)&=&\frac{\varepsilon^{ijk}\varepsilon^{imn}}{\sqrt{2}}\Big[u^{Tj}(x)C\gamma_5 c^k(x)  \bar{d}^m(x)\sigma_{\mu\nu} C \bar{c}^{Tn}(x)+ u^{Tj}(x)C\sigma_{\mu\nu} c^k(x)  \bar{d}^m(x)\gamma_5 C \bar{c}^{Tn}(x) \Big] \, , \nonumber\\
J^{P\widetilde{A}}_{-,\mu\nu}(x)&=&\frac{\varepsilon^{ijk}\varepsilon^{imn}}{\sqrt{2}}\Big[u^{Tj}(x)C c^k(x)  \bar{d}^m(x)\gamma_5\sigma_{\mu\nu} C \bar{c}^{Tn}(x)- u^{Tj}(x)C\sigma_{\mu\nu}\gamma_5 c^k(x)  \bar{d}^m(x)  C \bar{c}^{Tn}(x) \Big] \, , \nonumber\\
J^{P\widetilde{A}}_{+,\mu\nu}(x)&=&\frac{\varepsilon^{ijk}\varepsilon^{imn}}{\sqrt{2}}\Big[u^{Tj}(x)C c^k(x)  \bar{d}^m(x)\gamma_5\sigma_{\mu\nu} C \bar{c}^{Tn}(x)+ u^{Tj}(x)C\sigma_{\mu\nu}\gamma_5 c^k(x)  \bar{d}^m(x)  C \bar{c}^{Tn}(x) \Big] \, , \nonumber\\
\end{eqnarray}

\begin{eqnarray}
J^{AA}_{-,\mu\nu}(x)&=&\frac{\varepsilon^{ijk}\varepsilon^{imn}}{\sqrt{2}}\Big[u^{Tj}(x) C\gamma_\mu c^k(x) \bar{d}^m(x) \gamma_\nu C \bar{c}^{Tn}(x)  -u^{Tj}(x) C\gamma_\nu c^k(x) \bar{d}^m(x) \gamma_\mu C \bar{c}^{Tn}(x) \Big] \, ,  \nonumber\\
\end{eqnarray}
 the $i$, $j$, $k$, $m$, $n$ are  color indexes,   the $C$ is the charge conjugation matrix, the subscripts $\pm$ stand for the positive  and negative charge-conjugation, respectively, the superscripts $P$, $S$, $V$($\widetilde{V}$) and $A$($\widetilde{A}$) stand for  the pseudoscalar, scalar, vector and axialvector diquark and antidiquark operators, respectively.   The current $J^{PA}_{-,\mu}(x)$ was studied in Refs.\cite{WangY4360Y4660-1803,Wang-tetra-formula}, the current $J^{PA}_{+,\mu}(x)$ was studied in Ref.\cite{Wang-tetra-formula},   the current $J^{AA}_{-,\mu\nu}(x)$ was studied in Ref.\cite{WangEPJC-1601-Mc}, while the current $J^{S\widetilde{V}}_{-,\mu\nu}(x)$ was studied in Ref.\cite{Vector-Tetra-WZG-4100}. In the present work, we update the old calculations and perform new analysis in a comprehensive way,  and  make great efforts to exhaust all the possible vector tetraquark configurations without introducing an explicit P-wave. There are thirteen currents, we acquire original results for the eight currents, $J^{SV}_{+,\mu}(x)$,
$J_{-,\mu}^{\widetilde{V}A}(x)$,
$J_{+,\mu}^{\widetilde{V}A}(x)$,
 $J_{-,\mu}^{\widetilde{A}V}(x)$,
 $J_{+,\mu}^{\widetilde{A}V}(x)$,
$J^{S\widetilde{V}}_{+,\mu\nu}(x)$,
$J^{P\widetilde{A}}_{-,\mu\nu}(x)$ and
$J^{P\widetilde{A}}_{+,\mu\nu}(x)$.

  Under the parity transform $\widehat{P}$, the four-quark current  operators $J_\mu(x)$ and $J_{\mu\nu}(x)$ have the  properties,
\begin{eqnarray}\label{J-parity}
\widehat{P} J_\mu(x)\widehat{P}^{-1}&=&+J^\mu(\tilde{x}) \, , \nonumber\\
\widehat{P} \tilde{J}_{\mu\nu}(x)\widehat{P}^{-1}&=&-\tilde{J}^{\mu\nu}(\tilde{x}) \, , \nonumber\\
\widehat{P} J^{AA}_{-,\mu\nu}(x)\widehat{P}^{-1}&=&+J_{AA}^{-,\mu\nu}(\tilde{x}) \, ,
\end{eqnarray}
according to the properties of the diquark constituents,
\begin{eqnarray}
\widehat{P} \varepsilon^{ijk}q^T_j(x) C\Gamma c_k(x)\widehat{P}^{-1}&=&-\varepsilon^{ijk}q^T_j(\tilde{x}) C\gamma^0\Gamma\gamma^0 c_k(\tilde{x}) \, , \nonumber\\
\widehat{P} \varepsilon^{ijk}\bar{q}_j(x)\Gamma C \bar{c}^T_k(x)\widehat{P}^{-1}&=&-\varepsilon^{ijk}\bar{q}_j(\tilde{x}) \gamma^0\Gamma\gamma^0 C\bar{c}^T_k(\tilde{x}) \, ,
\end{eqnarray}
where $\tilde{J}_{\mu\nu}(x)=J^{S\widetilde{V}}_{-,\mu\nu}(x)$,
$J^{S\widetilde{V}}_{+,\mu\nu}(x)$,
$J^{P\widetilde{A}}_{-,\mu\nu}(x)$,
$J^{P\widetilde{A}}_{+,\mu\nu}(x)$, the coordinates $x^\mu=(t,\vec{x})$ and $\tilde{x}^\mu=(t,-\vec{x})$. For $\Gamma=1$, $\gamma_5$, $\gamma_\mu$,  $\gamma_\mu\gamma_5$, $\sigma_{\mu\nu}$, $\sigma_{\mu\nu}\gamma_5$, we acquire $\gamma^0\Gamma\gamma^0=1$, $-\gamma_5$, $\gamma^\mu$,  $-\gamma^\mu\gamma_5$, $\sigma^{\mu\nu}$, $-\sigma^{\mu\nu}\gamma_5$.
 We rewrite Eq.\eqref{J-parity} in  more explicit form,
\begin{eqnarray}
\widehat{P} J_i(x)\widehat{P}^{-1}&=&-J_i(\tilde{x}) \, , \nonumber\\
\widehat{P} \tilde{J}_{ij}(x)\widehat{P}^{-1}&=&-\tilde{J}_{ij}(\tilde{x}) \, , \nonumber\\
\widehat{P} J^{AA}_{-,0i}(x)\widehat{P}^{-1}&=&-J_{AA,0i}^{-}(\tilde{x}) \, ,
\end{eqnarray}
\begin{eqnarray}
\widehat{P} J_0(x)\widehat{P}^{-1}&=&+J_0(\tilde{x}) \, , \nonumber\\
\widehat{P} \tilde{J}_{0i}(x)\widehat{P}^{-1}&=&+\tilde{J}_{0i}(\tilde{x}) \, , \nonumber\\
\widehat{P} J^{AA}_{-,ij}(x)\widehat{P}^{-1}&=&+J_{AA,ij}^{-}(\tilde{x}) \, ,
\end{eqnarray}
where $i$, $j=1$, $2$, $3$. Now we can see clearly that the currents $J_\mu(x)$ and $J_{\mu\nu}(x)$ have both  negative-parity and positive-parity components, which couple potentially to the hidden-charm tetraquark states with the  negative-parity and positive-parity, respectively, we separate their contributions explicitly by introducing the suitable projectors in practical calculations.

Under the charge-conjugation transform $\widehat{C}$, the four-quark currents  $J_\mu(x)$ and $J_{\mu\nu}(x)$ have the properties,
\begin{eqnarray}
\widehat{C}J_{\pm,\mu}(x)\widehat{C}^{-1}&=&\pm J_{\pm,\mu}(x)\mid_{u\leftrightarrow d}  \, , \nonumber\\
\widehat{C}J_{\pm,\mu\nu}(x)\widehat{C}^{-1}&=&\pm J_{\pm,\mu\nu}(x)\mid_{u\leftrightarrow d}  \, .
\end{eqnarray}

 The four-quark currents  $J_\mu(x)$ and $J_{\mu\nu}(x)$ with hidden-charm have the  symbolic  structure  $\bar{c}c\bar{d}u$ and the isospin $(I,I_3)=(1,1)$,
 we can construct other currents in the isospin multiplets in a similar way.
 In the isospin limit $m_u=m_d=m_q$, the four-quark currents with the  symbolic  structures,
 \begin{eqnarray}
 \bar{c}c\bar{d}u, \, \, \bar{c}c\bar{u}d, \, \, \bar{c}c\frac{\bar{u}u-\bar{d}d}{\sqrt{2}}, \, \, \bar{c}c\frac{\bar{u}u+\bar{d}d}{\sqrt{2}}\, ,
 \end{eqnarray}
 couple potentially  to the vector hidden-charm
tetraquark states with almost degenerated  masses, the currents with the isospin $I=1$ and $0$ result in the same expressions of the QCD sum rules.
Only the four-quark currents with the symbolic structures $\bar{c}c\frac{\bar{u}u-\bar{d}d}{\sqrt{2}}$ and $\bar{c}c\frac{\bar{u}u+\bar{d}d}{\sqrt{2}}$ have definite
charge-conjugation, and we take it for granted  that the vector  tetraquark states $\bar{c}c\bar{d}u$ have the same charge-conjugation as their charge-neutral  cousins.

At the  phenomenological  side, we  insert  a perfect  set of intermediate hadronic states with
the same quantum numbers (spin, parity, charge-conjugation) as the local four-quark currents  $J_\mu(x)$ and $J_{\mu\nu}(x)$ into the
correlation functions  $\Pi_{\mu\nu}(p)$ and $\Pi_{\mu\nu\alpha\beta}(p)$   to acquire the hadronic spectral representation
\cite{SVZ79-1,SVZ79-2,Reinders85}, and separate  the contributions of the lowest vector hidden-charm tetraquark states,
\begin{eqnarray}
\Pi_{\mu\nu}(p)&=&\frac{\lambda_{Y_{-}}^2}{M_{Y_{-}}^2-p^2}\left( -g_{\mu\nu}+\frac{p_{\mu}p_{\nu}}{p^2}\right) +\cdots \nonumber\\
&=&\Pi_{-}(p^2)\left( -g_{\mu\nu}+\frac{p_{\mu}p_{\nu}}{p^2}\right)+\cdots \, ,\nonumber\\
\widetilde{\Pi}_{\mu\nu\alpha\beta}(p)&=&\frac{\lambda_{Y_{-}}^2}{M_{Y_{-}}^2\left(M_{Y_{-}}^2-p^2\right)}\left(p^2g_{\mu\alpha}g_{\nu\beta} -p^2g_{\mu\beta}g_{\nu\alpha} -g_{\mu\alpha}p_{\nu}p_{\beta}-g_{\nu\beta}p_{\mu}p_{\alpha}+g_{\mu\beta}p_{\nu}p_{\alpha}+g_{\nu\alpha}p_{\mu}p_{\beta}\right) \nonumber\\
&&+\frac{\lambda_{ Z_{+}}^2}{M_{Z_{+}}^2\left(M_{Z_{+}}^2-p^2\right)}\left( -g_{\mu\alpha}p_{\nu}p_{\beta}-g_{\nu\beta}p_{\mu}p_{\alpha}+g_{\mu\beta}p_{\nu}p_{\alpha}+g_{\nu\alpha}p_{\mu}p_{\beta}\right) +\cdots  \nonumber\\
&=&\widetilde{\Pi}_{-}(p^2)\left(p^2g_{\mu\alpha}g_{\nu\beta} -p^2g_{\mu\beta}g_{\nu\alpha} -g_{\mu\alpha}p_{\nu}p_{\beta}-g_{\nu\beta}p_{\mu}p_{\alpha}+g_{\mu\beta}p_{\nu}p_{\alpha}+g_{\nu\alpha}p_{\mu}p_{\beta}\right) \nonumber\\
&&+\widetilde{\Pi}_{+}(p^2)\left( -g_{\mu\alpha}p_{\nu}p_{\beta}-g_{\nu\beta}p_{\mu}p_{\alpha}+g_{\mu\beta}p_{\nu}p_{\alpha}+g_{\nu\alpha}p_{\mu}p_{\beta}\right) \, ,\nonumber\\
\Pi^{AA}_{\mu\nu\alpha\beta}(p)&=&\frac{\lambda_{ Z_{+}}^2}{M_{Z_{+}}^2\left(M_{Z_{+}}^2-p^2\right)}\left(p^2g_{\mu\alpha}g_{\nu\beta} -p^2g_{\mu\beta}g_{\nu\alpha} -g_{\mu\alpha}p_{\nu}p_{\beta}-g_{\nu\beta}p_{\mu}p_{\alpha}+g_{\mu\beta}p_{\nu}p_{\alpha}+g_{\nu\alpha}p_{\mu}p_{\beta}\right) \nonumber\\
&&+\frac{\lambda_{Y_{ -}}^2}{M_{Y_{-}}^2\left(M_{Y_{-}}^2-p^2\right)}\left( -g_{\mu\alpha}p_{\nu}p_{\beta}-g_{\nu\beta}p_{\mu}p_{\alpha}+g_{\mu\beta}p_{\nu}p_{\alpha}+g_{\nu\alpha}p_{\mu}p_{\beta}\right) +\cdots  \nonumber\\
&=&\widetilde{\Pi}_{+}(p^2)\left(p^2g_{\mu\alpha}g_{\nu\beta} -p^2g_{\mu\beta}g_{\nu\alpha} -g_{\mu\alpha}p_{\nu}p_{\beta}-g_{\nu\beta}p_{\mu}p_{\alpha}+g_{\mu\beta}p_{\nu}p_{\alpha}+g_{\nu\alpha}p_{\mu}p_{\beta}\right) \nonumber\\
&&+\widetilde{\Pi}_{-}(p^2)\left( -g_{\mu\alpha}p_{\nu}p_{\beta}-g_{\nu\beta}p_{\mu}p_{\alpha}+g_{\mu\beta}p_{\nu}p_{\alpha}+g_{\nu\alpha}p_{\mu}p_{\beta}\right) \, ,
\end{eqnarray}
where the pole residues $\lambda_Y$ and $\lambda_Z$ are defined by
\begin{eqnarray}
  \langle 0|J_\mu(0)|Y_c^-(p)\rangle &=&\lambda_{Y_{-}}\varepsilon_\mu\, , \nonumber\\
  \langle 0|\tilde{J}_{\mu\nu}(0)|Y_c^-(p)\rangle &=& \frac{\lambda_{Y_{-}}}{M_{Y_{-}}} \, \varepsilon_{\mu\nu\alpha\beta} \, \varepsilon^{\alpha}p^{\beta}\, , \nonumber\\
 \langle 0|\tilde{J}_{\mu\nu}(0)|Z_c^+(p)\rangle &=&\frac{\lambda_{Z_{+}}}{M_{Z_{+}}} \left(\varepsilon_{\mu}p_{\nu}-\varepsilon_{\nu}p_{\mu} \right)\, , \nonumber\\
  \langle 0|J_{-,\mu\nu}^{AA}(0)|Z_c^+(p)\rangle &=& \frac{\lambda_{Z_{+}}}{M_{Z_{+}}} \, \varepsilon_{\mu\nu\alpha\beta} \, \varepsilon^{\alpha}p^{\beta}\, , \nonumber\\
 \langle 0|J_{-,\mu\nu}^{AA}(0)|Y_c^-(p)\rangle &=&\frac{\lambda_{Y_{-}}}{M_{Y_{-}}} \left(\varepsilon_{\mu}p_{\nu}-\varepsilon_{\nu}p_{\mu} \right)\, ,
\end{eqnarray}
the  $\varepsilon_{\mu/\alpha}$ and $\varepsilon_{\mu\nu}$ are the polarization vectors, we add the superscripts/subscripts $\pm$  to indicate the positive and negative parity, respectively. We choose the components $\Pi_{-}(p^2)$ and $p^2\widetilde{\Pi}_{-}(p^2)$ to explore the negative-parity hidden-charm tetraquark states with the angular momentum $J=1$.

At the QCD side of the correlation functions $\Pi_{\mu\nu}(p)$ and $\Pi_{\mu\nu\alpha\beta}(p)$, there are  two heavy  quark propagators and two light quark propagators.  Supposing that each heavy quark line emits a gluon and each light quark line contributes  a quark-antiquark  pair, we acquire a quark-gluon  operator $g_sGg_sG\bar{u}u \bar{d}d$   of dimension 10, therefore we should calculate  the vacuum condensates at least
up to dimension 10 to evaluate  the convergent behavior of the operator product expansion, and take into account the vacuum condensates $\langle\bar{q}q\rangle$, $\langle\frac{\alpha_{s}GG}{\pi}\rangle$, $\langle\bar{q}g_{s}\sigma Gq\rangle$, $\langle\bar{q}q\rangle^2$, $g_s^2\langle\bar{q}q\rangle^2$,
$\langle\bar{q}q\rangle \langle\frac{\alpha_{s}GG}{\pi}\rangle$,  $\langle\bar{q}q\rangle  \langle\bar{q}g_{s}\sigma Gq\rangle$,
$\langle\bar{q}g_{s}\sigma Gq\rangle^2$ and $\langle\bar{q}q\rangle^2 \langle\frac{\alpha_{s}GG}{\pi}\rangle$, which are vacuum expectations of the quark-gluon operators of the order $\mathcal{O}(\alpha_s^k)$ with $k\leq 1$ \cite{Wang-tetra-formula,WangHuangtao-2014-PRD,WZG-mole-EPJC-1}.  The highest  vacuum condensates  $\langle\bar{q}g_{s}\sigma Gq\rangle^2$ and $\langle\bar{q}q\rangle^2 \langle\frac{\alpha_{s}GG}{\pi}\rangle$ are associated with $\frac{1}{T^2}$, $\frac{1}{T^4}$ and  $\frac{1}{T^6}$, and  play an important role in acquiring  the Borel windows, although  they are of minor importance or play a tiny role in the Borel windows. We  recalculate the contributions having  their origins in the higher  dimensional vacuum condensates according to  the identity $t^a_{ij}t^a_{mn}=-\frac{1}{6}\delta_{ij}\delta_{mn}+\frac{1}{2}\delta_{jm}\delta_{in}$, and acquire slightly  different analytical expressions  compared with
  the old calculations for the three currents $J^{PA}_{-,\mu}(x)$, $J^{PA}_{+,\mu}(x)$ and $J^{AA}_{-,\mu\nu}(x)$ \cite{WangY4360Y4660-1803,Wang-tetra-formula,WangEPJC-1601-Mc}, where $t^a=\frac{\lambda^a}{2}$,  the $\lambda^a$ is the Gell-Mann matrix.
  The four-quark condensate $g_s^2\langle \bar{q}q\rangle^2$ has  its origins in the vacuum expectations
$\langle \bar{q}\gamma_\mu t^a q g_s D_\eta G^a_{\lambda\tau}\rangle$, $\langle\bar{q}_jD^{\dagger}_{\mu}D^{\dagger}_{\nu}D^{\dagger}_{\alpha}q_i\rangle$  and
$\langle\bar{q}_jD_{\mu}D_{\nu}D_{\alpha}q_i\rangle$ combined with the vacuum saturation assumption, rather than  has  its origins in the radiative  corrections for the four-quark condensate $\langle \bar{q}q\rangle^2$, where $D_\alpha=\partial_\alpha-ig_sG^a_\alpha t^a$,  its contributions are tiny and neglected in most of the QCD sum rules.  The strong fine-structure-constant/coupling-constant $\alpha_s(\mu)=\frac{g_s^2(\mu)}{4\pi}$ appears  at the tree level, which is energy scale dependent and manifests the  necessity  of applying the energy scale formula $\mu=\sqrt{M^2_{X/Y/Z}-(2{\mathbb{M}}_c)^2}$.

  We match  the phenomenological side with the QCD  side of the components $\Pi_{-}(p^2)$ and $p^2\widetilde{\Pi}_{-}(p^2)$ below the continuum thresholds   $s_0$ with the help of spectral representation, and perform Borel transform  in regard  to the variable
 $P^2=-p^2$ to acquire   the  QCD sum rules:
\begin{eqnarray}\label{QCDSR}
\lambda^2_{Y}\, \exp\left(-\frac{M^2_{Y}}{T^2}\right)= \int_{4m_c^2}^{s_0} ds\, \rho(s) \, \exp\left(-\frac{s}{T^2}\right) \, ,
\end{eqnarray}
the  explicit expressions of the QCD spectral densities $\rho(s)$ are neglected for simplicity.

We differentiate  Eq.\eqref{QCDSR} in regard to the variable  $\tau=\frac{1}{T^2}$,  and acquire the QCD sum rules for
 the masses of the  vector hidden-charm tetraquark states $Y_c$,
 \begin{eqnarray}\label{mass-QCDSR}
 M^2_{Y}&=& -\frac{\int_{4m_c^2}^{s_0} ds\frac{d}{d \tau}\rho(s)\exp\left(-\tau s \right)}{\int_{4m_c^2}^{s_0} ds \rho(s)\exp\left(-\tau s\right)}\, .
\end{eqnarray}

\section{Numerical results and discussions}
The quark masses and vacuum condensates depend on the energy scale, we write down the energy-scale dependence of  the input parameters at the QCD side,
\begin{eqnarray}
\langle\bar{q}q \rangle(\mu)&=&\langle\bar{q}q \rangle({\rm 1GeV})\left[\frac{\alpha_{s}({\rm 1GeV})}{\alpha_{s}(\mu)}\right]^{\frac{12}{33-2n_f}}\, , \nonumber\\
 \langle\bar{q}g_s \sigma Gq \rangle(\mu)&=&\langle\bar{q}g_s \sigma Gq \rangle({\rm 1GeV})\left[\frac{\alpha_{s}({\rm 1GeV})}{\alpha_{s}(\mu)}\right]^{\frac{2}{33-2n_f}}\, , \nonumber\\
 m_c(\mu)&=&m_c(m_c)\left[\frac{\alpha_{s}(\mu)}{\alpha_{s}(m_c)}\right]^{\frac{12}{33-2n_f}} \, ,\nonumber\\
\alpha_s(\mu)&=&\frac{1}{b_0t}\left[1-\frac{b_1}{b_0^2}\frac{\log t}{t} +\frac{b_1^2(\log^2{t}-\log{t}-1)+b_0b_2}{b_0^4t^2}\right]\, ,
\end{eqnarray}
 from the renormalization group equation,  where $t=\log \frac{\mu^2}{\Lambda_{QCD}^2}$, $b_0=\frac{33-2n_f}{12\pi}$, $b_1=\frac{153-19n_f}{24\pi^2}$, $b_2=\frac{2857-\frac{5033}{9}n_f+\frac{325}{27}n_f^2}{128\pi^3}$,  $\Lambda_{QCD}=210\,\rm{MeV}$, $292\,\rm{MeV}$  and  $332\,\rm{MeV}$ for the flavors  $n_f=5$, $4$ and $3$, respectively  \cite{PDG,Narison-mix}. In this work, as the up, down and charm quarks are concerned, we adopt the flavor number $n_f=4$ and neglect the small $u$ and $d$ quark masses.

 At the beginning points, we adopt  the standard-values/conventional-values of the vacuum condensates $\langle
\bar{q}q \rangle=-(0.24\pm 0.01\, \rm{GeV})^3$,   $\langle
\bar{q}g_s\sigma G q \rangle=m_0^2\langle \bar{q}q \rangle$,
$m_0^2=(0.8 \pm 0.1)\,\rm{GeV}^2$,  $\langle \frac{\alpha_s
GG}{\pi}\rangle=(0.012\pm0.004)\,\rm{GeV}^4 $    at the typical  energy scale  $\mu=1\, \rm{GeV}$
\cite{SVZ79-1,SVZ79-2,Reinders85,Colangelo-Review}, and adopt the modified minimal subtracted mass $m_{c}(m_c)=(1.275\pm0.025)\,\rm{GeV}$ from the Particle Data Group \cite{PDG}.

In this work, we resort to the energy scale formula $\mu=\sqrt{M^2_{X/Y/Z}-(2{\mathbb{M}}_c)^2}$ with the updated value  ${\mathbb{M}}_c=1.82\,\rm{GeV}$ to determine the pertinent energy scales of the QCD spectral densities in the QCD sum rules \cite{Wang-tetra-formula,WangEPJC-1601-Mc}.
We can rewrite the energy scale formula in the form,
\begin{eqnarray}\label{formula-Regge}
M^2_{X/Y/Z}&=&\mu^2+{\rm Constants}\, ,
\end{eqnarray}
where the Constants have the value $4{\mathbb{M}}_c^2$ and are fitted phenomenologically by the QCD sum rules, the predicted tetraquark  masses and the pertinent  energy scales of the QCD spectral densities have a  Regge-trajectory-like relation. The constraint shown in Eq.\eqref{formula-Regge}  plays an important role in enhancing  the pole contributions and improving the convergent behaviors of the operator product expansion so as to acquire the lowest  tetraquark  (molecular) masses, which are calculated  with the QCD sum rules in Eq.\eqref{mass-QCDSR}.

Now let us take a short digression to discuss why the QCD sum rules for the hidden-charm tetraquark (molecular)  states depend on the energy scales. We can write the correlation functions $\Pi(p^2)$ for the hidden-charm four-quark currents  in the form,
\begin{eqnarray}
\Pi(p^2)&=&\int_{4m^2_c(\mu)}^{s_0} ds \frac{\rho_{QCD}(s,\mu)}{s-p^2}+\int_{s_0}^\infty ds \frac{\rho_{QCD}(s,\mu)}{s-p^2} \, ,
\end{eqnarray}
through dispersion relation, and they are energy scale independent or independent on the energy scale we select to carry out the operator product expansion (up to a factor of the anomalous dimensions of the currents $\gamma_J$), but which does not ensure the contributions of the ground states,
\begin{eqnarray}
\frac{d}{d\mu}\int_{4m^2_c(\mu)}^{s_0} ds \frac{\rho_{QCD}(s,\mu)}{s-p^2}\rightarrow 0 \, .
\end{eqnarray}
 In carrying out the operator product expansion, we usually neglect the radiative  corrections, even in the QCD sum rules for the traditional/conventional mesons, we fail to  take account of the radiative  corrections up to arbitrary  orders; we factorize the higher dimensional vacuum condensates  into lower dimensional ones according to vacuum saturation, consequently we modify  the energy scale dependence of the higher dimensional vacuum condensates  more or less.
Furthermore, we introduce the truncations $s_0$ to exclude the contaminations from  the continuum states and higher resonances, and we have no knowledge of the possible correlation between the threshold $4m^2_c(\mu)$ and continuum threshold $s_0$.
After performing the Borel transform, we acquire the integrals,
 \begin{eqnarray}
 \int_{4m_c^2(\mu)}^{s_0} ds \rho_{QCD}(s,\mu)\exp\left(-\frac{s}{T^2} \right)\, ,
 \end{eqnarray}
which are sensitive to  the energy scale $\mu$.  Variations of the energy scale $\mu$ can result in variations of the integral ranges $4m_c^2(\mu)-s_0$ of the variable
$ds$ besides the QCD spectral densities $\rho_{QCD}(s,\mu)$, consequently, they result in variations of the Borel windows and predicted tetraquark (molecule) masses and pole residues. We resort to the energy scale formula to determine the pertinent energy scales consistently.

We usually consult the experimental data on  the mass gaps between the ground states (1S) and first radial excited states (2S)  to acquire the continuum threshold parameters $s_0$.
According to the (possible) quantum numbers (such as spin, parity, charge-conjugation, etc), decay channels and mass  gaps, if we prefer the scenarios of tetraquark states to other interpretations, we can tentatively identify the $X(3915)$ and $X(4500)$ as the 1S and 2S  hidden-charm  tetraquark states with the quantum numbers $J^{PC}=0^{++}$ \cite{X4140-tetraquark-Lebed,X3915-X4500-EPJC-WZG}, identify
the $Z_c(3900)$ and $Z_c(4430)$   as  the 1S and 2S hidden-charm tetraquark states with the quantum numbers $J^{PC}=1^{+-}$, respectively \cite{Maiani-II-type,Nielsen-1401,WangZG-Z4430-CTP},   identify the $Z_c(4020)$ and $Z_c(4600)$ as the 1S and 2S hidden-charm tetraquark states with the quantum numbers $J^{PC}=1^{+-}$, respectively  \cite{ChenHX-Z4600-A,WangZG-axial-Z4600}, and identify the $X(4140)$ and $X(4685)$ as the 1S and 2S hidden-charm tetraquark states with the quantum numbers $J^{PC}=1^{++}$, respectively \cite{WZG-X4140-X4685}. The mass gaps between the 1S and 2S hidden-charm tetraquark states  are about $0.57\sim 0.59 \,\rm{GeV}$. In this work, we can set the continuum threshold parameters as  $\sqrt{s_0}=M_Y+0.4\sim0.6\,\rm{GeV}$. Compared with previous works \cite{WangY4360Y4660-1803,Wang-tetra-formula,WangEPJC-1601-Mc,Vector-Tetra-WZG-4100}, we adopt  the uniform constraint between the ground state masses $M_Y$ and continuum threshold parameters $s_0$, $\sqrt{s_0}=M_Y+0.5\pm 0.1\,\rm{GeV}$, for all the vector tetraquark states, and perform a consistent analysis by including additional eight novel currents (or QCD sum rules). Furthermore, we correct some minor errors in the numerical calculations in Ref.\cite{Wang-tetra-formula}.

The pole or ground state dominance at the phenomenological  side and convergence of the operator product expansion at the QCD side are two elementary   criteria, we should satisfy the two elementary criteria to acquire reliable/robust QCD sum rules.
Now we write down explicitly the definitions for the  pole contributions (PC),
\begin{eqnarray}
{\rm{PC}}&=&\frac{\int_{4m_{c}^{2}}^{s_{0}}ds\rho\left(s\right)\exp\left(-\frac{s}{T^{2}}\right)} {\int_{4m_{c}^{2}}^{\infty}ds\rho\left(s\right)\exp\left(-\frac{s}{T^{2}}\right)}\, ,
\end{eqnarray}
 and the contributions of the vacuum condensates  of dimension $n$,
\begin{eqnarray}
D(n)&=&\frac{\int_{4m_{c}^{2}}^{s_{0}}ds\rho_{n}(s)\exp\left(-\frac{s}{T^{2}}\right)}
{\int_{4m_{c}^{2}}^{s_{0}}ds\rho\left(s\right)\exp\left(-\frac{s}{T^{2}}\right)}\, .
\end{eqnarray}
 In the present work, we require  the contributions $|D(10)|\sim 1\%$ or $< 1\%$  at the Borel windows.

\begin{figure}
 \centering
 \includegraphics[totalheight=6cm,width=7cm]{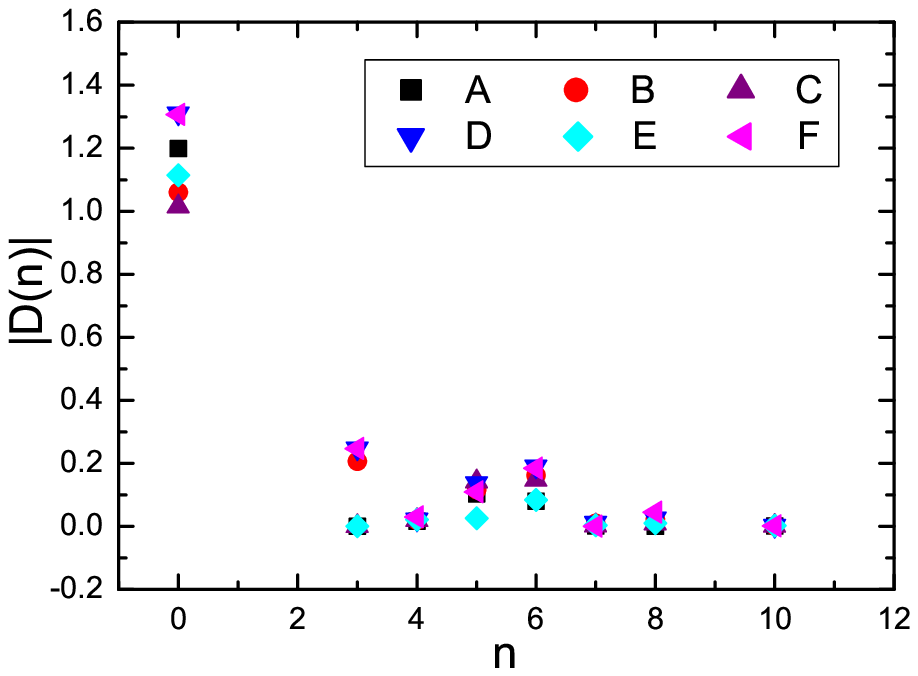}
 \includegraphics[totalheight=6cm,width=7cm]{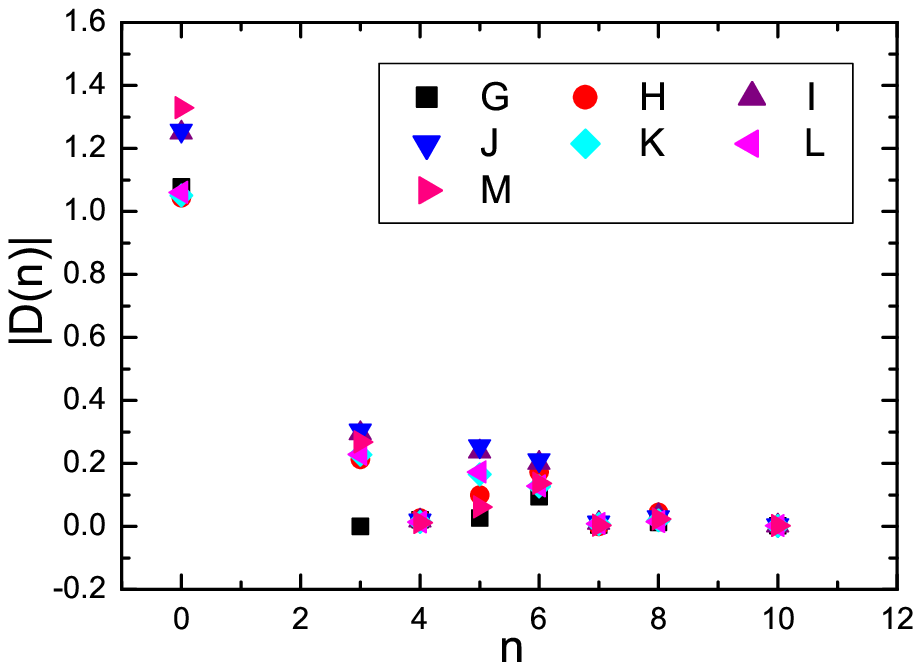}
 \caption{ The absolute contributions of the vacuum condensates with the central values of the input parameters, where the $A$, $B$, $C$, $D$, $E$, $F$, $G$, $H$, $I$, $J$, $K$, $L$ and $M$ stand for the $[uc]_{P}[\overline{dc}]_{A}-[uc]_{A}[\overline{dc}]_{P}$,
$[uc]_{P}[\overline{dc}]_{A}+[uc]_{A}[\overline{dc}]_{P}$, $[uc]_{S}[\overline{dc}]_{V}+[uc]_{V}[\overline{dc}]_{S}$,
$[uc]_{S}[\overline{dc}]_{V}-[uc]_{V}[\overline{dc}]_{S}$, $[uc]_{\tilde{V}}[\overline{dc}]_{A}-[uc]_{A}[\overline{dc}]_{\tilde{V}}$,
$[uc]_{\tilde{V}}[\overline{dc}]_{A}+[uc]_{A}[\overline{dc}]_{\tilde{V}}$, $[uc]_{\tilde{A}}[\overline{dc}]_{V}+[uc]_{V}[\overline{dc}]_{\tilde{A}}$,
$[uc]_{\tilde{A}}[\overline{dc}]_{V}-[uc]_{V}[\overline{dc}]_{\tilde{A}}$, $[uc]_{S}[\overline{dc}]_{\tilde{V}}-[uc]_{\tilde{V}}[\overline{dc}]_{S}$,
$[uc]_{S}[\overline{dc}]_{\tilde{V}}+[uc]_{\tilde{V}}[\overline{dc}]_{S}$, $[uc]_{P}[\overline{dc}]_{\tilde{A}}-[uc]_{\tilde{A}}[\overline{dc}]_{P}$,
$[uc]_{P}[\overline{dc}]_{\tilde{A}}+[uc]_{\tilde{A}}[\overline{dc}]_{P}$ and $[uc]_{A}[\overline{dc}]_{A}$ vector tetraquark states, respectively.                                          }\label{OPE-fig}
\end{figure}

 We search for the best Borel parameters and continuum threshold parameters via trial and error. At last,  we acquire the Borel windows, continuum threshold parameters, suitable energy scales of the QCD spectral densities and  pole contributions, which are shown explicitly in Table \ref{BorelP}.
From the Table,  we can see plainly that the pole contributions are about $(40-60)\%$ at the phenomenological side, while the central values are larger than $50\%$, the pole dominance criterion  is  satisfied very good.
In Fig.\ref{OPE-fig}, we plot the absolute contributions of the vacuum condensates $|D(n)|$ with the central values of the input parameters shown in Table \ref{BorelP}. From the figure, we can see clearly that the main contributions have their origins in the perturbative terms, the higher dimensional vacuum condensates play a minor important role (or they are of tiny importance).  For example, the contributions of the vacuum condensates of dimension $10$ are $|D(10)|\ll 1\%$, just like our expectation,  the convergent behavior of the operator product  expansion is very good.

\begin{figure}
 \centering
 \includegraphics[totalheight=6cm,width=7cm]{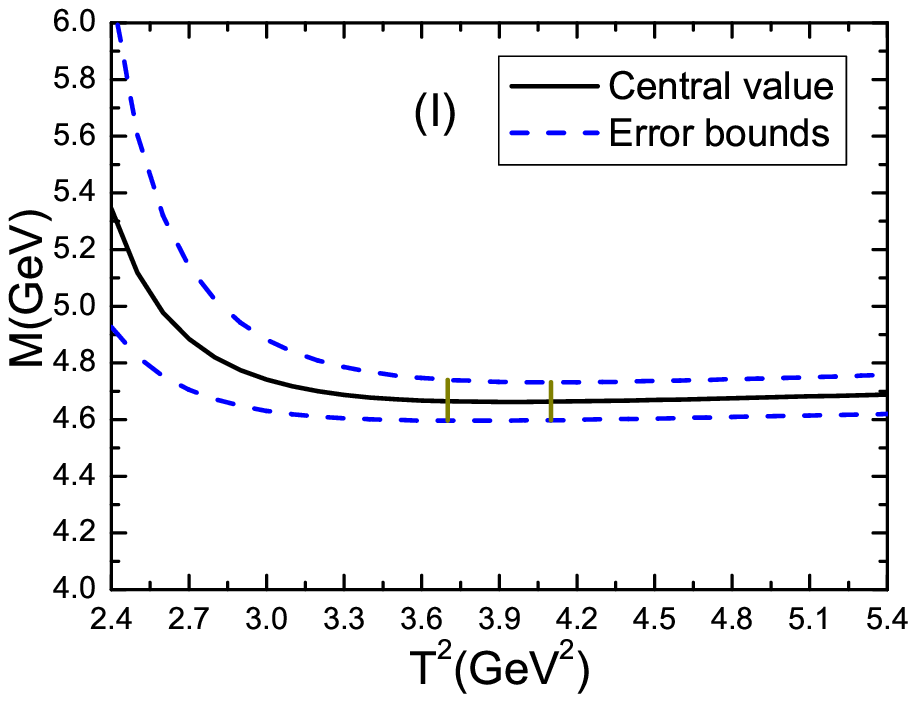}
 \includegraphics[totalheight=6cm,width=7cm]{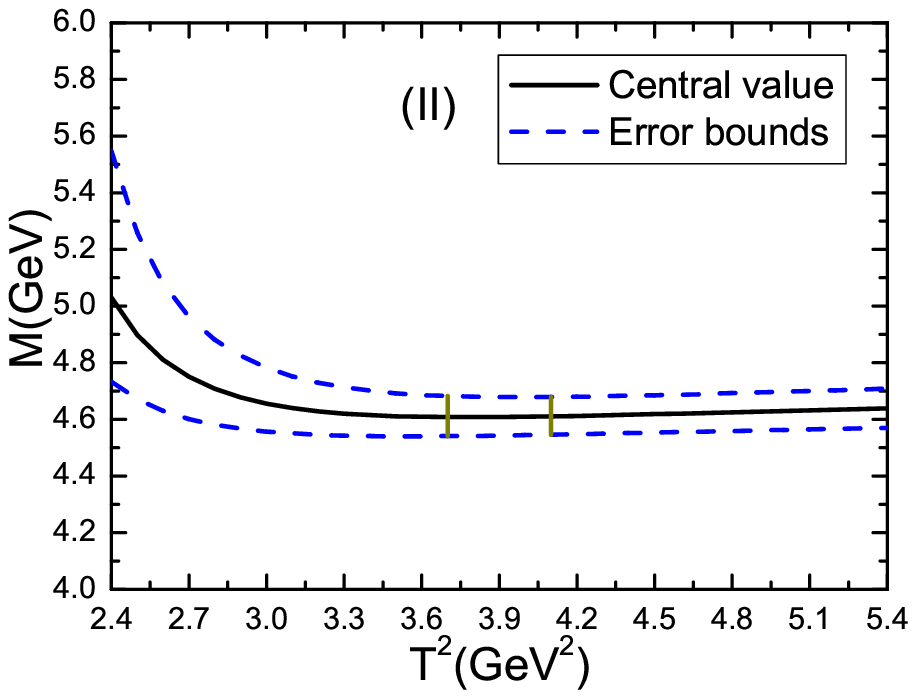}
 \caption{ The masses of the  $[uc]_P[\overline{dc}]_{A}-[uc]_{A}[\overline{dc}]_P$(I) and $[uc]_P[\overline{dc}]_{A}+[uc]_{A}[\overline{dc}]_P$(II) vector tetraquark states   via  variations  of the Borel parameters $T^2$, where the regions between the two vertical lines are the Borel windows.   }\label{mass-1-fig}
\end{figure}

We take account of all the uncertainties from  the relevant parameters and acquire the masses and pole residues of the vector hidden-charm  tetraquark states with both the positive and negative charge-conjugation from the coupled equations \eqref{QCDSR} and \eqref{mass-QCDSR}, and we also present them clearly in Table \ref{BorelP}. From  Table \ref{BorelP}, we can see clearly that the energy scale formula $\mu=\sqrt{M^2_{X/Y/Z}-(2{\mathbb{M}}_c)^2}$ is  satisfied very good. The energy scale formula plays an important role (or is of  crucial importance) in enhancing  the pole contributions and improving the convergent behaviors of the operator product expansion in the QCD sum rules for the hidden-charm (or hidden-bottom) and doubly charmed (or doubly bottom) tetraquark and pentaquark (molecular) states \cite{Wang-tetra-formula,Vector-mole-WZG-CPC,WZG-HC-spectrum-PRD,WZG-Zcs-spectrum-CPC,WZG-mole-EPJC-1,WangZG-Z4430-CTP}.
 In  Fig.\ref{mass-1-fig}, we plot the predicted masses of the  $[uc]_P[\overline{dc}]_{A}-[uc]_{A}[\overline{dc}]_P$ and $[uc]_P[\overline{dc}]_{A}+[uc]_{A}[\overline{dc}]_P$ tetraquark states with the quantum numbers $J^{PC}=1^{--}$ and $1^{-+}$ respectively via variations of the Borel parameters at much larger ranges than the Borel widows as a typical example. From the figure, we can see explicitly  that there appear very flat platforms in the Borel windows as a matter of fact, which can exclude  additional uncertainties coming   from the Borel parameters.

If we abandon the energy scale formula $\mu=\sqrt{M^2_{X/Y/Z}-(2{\mathbb{M}}_c)^2}$ and choose the particular energy scale $\mu=1\,\rm{GeV}$, we can perform the same procedure by searching for the best Borel parameters and continuum threshold parameters via trial and error. At last, we acquire  the Borel windows, continuum threshold parameters and  pole contributions, which are presented plainly in Table \ref{BorelP-1GeV}. From Table \ref{BorelP-1GeV}, we can see clearly that the pole contributions are about $(40-60)\%$, just like the corresponding ones shown in Table \ref{BorelP}, again the pole dominance criterion  is satisfied very good. At the QCD side, the dominant contributions have their origins in the perturbative terms, the operator product expansion converges very well. Again, we take  account of all the uncertainties from the relevant  parameters and reach the masses and pole residues of the vector hidden-charm  tetraquark states, which are also presented plainly in Table \ref{BorelP-1GeV}.

In Table \ref{Interpretations-Table}, we present the possible interpretations of the ground state vector hidden-charm tetraquark states, the present predictions support identifying the $Y(4360/4390)$ as the $[uc]_{S}[\overline{dc}]_{V}+[uc]_{V}[\overline{dc}]_{S}$ hidden-charm tetraquark state with the quantum numbers $J^{PC}=1^{--}$ and identifying the
$Y(4660)$ as the $[uc]_{P}[\overline{dc}]_{A}-[uc]_{A}[\overline{dc}]_{P}$  or  $[uc]_{A}[\overline{dc}]_{A}$ hidden-charm tetraquark state with the quantum numbers  $J^{PC}=1^{--}$.

In Table \ref{Interpretations-Table}, we also present the masses of the ground state vector hidden-charm tetraquark states extracted at the particular energy scale $\mu=1\,\rm{GeV}$, the predicted tetraquark masses $\widehat{M}_Y$ are much larger than the corresponding ones $M_Y$, which disfavors  identifying all the existing $Y$ states as the vector tetraquark states due to the large masses.

In Ref.\cite{Vector-Tetra-WZG-P-wave}, the predicted mass $M_{Y}=4.24\pm0.10\,\rm{GeV}$ for the $|0, 0; 0, 1; 1\rangle$ tetraquark  state  supports  identifying the $Y(4260/4220)$  as the  $C\gamma_5\otimes\stackrel{\leftrightarrow}{\partial}_\mu\otimes \gamma_5C$   type vector hidden-charm  tetraquark state;
 the predicted mass $M_{Y}=4.28\pm0.10\,\rm{GeV}$ for the $|1, 1; 0, 1; 1\rangle$ tetraquark state supports  identifying the $Y(4260/4220)$ or $Y(4360/4320)$ as the  $C\gamma_\alpha\otimes\stackrel{\leftrightarrow}{\partial}_\mu\otimes \gamma^{\alpha}C$   type vector hidden-charm tetraquark state;
 the predicted masses $M_{Y}=4.31\pm0.10\,\rm{GeV}$ for the $\frac{1}{\sqrt{2}}\left(|1, 0; 1, 1; 1\rangle+|0, 1; 1, 1; 1\rangle\right)$ tetraquark state  and  $M_{Y}=4.33\pm0.10\,\rm{GeV}$ for the $|1, 1; 2, 1; 1\rangle$ tetraquark state  supports  identifying the $Y(4360/4320)$ or $Y(4390)$ as the
 $C\gamma_\mu \otimes\stackrel{\leftrightarrow}{\partial}_\alpha \otimes\gamma^\alpha C +C\gamma^\alpha \otimes\stackrel{\leftrightarrow}{\partial}_\alpha \otimes\gamma_\mu C$ type or $C\gamma_5 \otimes\stackrel{\leftrightarrow}{\partial}_\mu \otimes\gamma_\nu C
+C\gamma_\nu \otimes\stackrel{\leftrightarrow}{\partial}_\mu \otimes\gamma_5 C
-C\gamma_5 \otimes\stackrel{\leftrightarrow}{\partial}_\nu \otimes\gamma_\mu C
-C\gamma_\mu \otimes\stackrel{\leftrightarrow}{\partial}_\nu \otimes\gamma_5 C $ type  vector hidden-charm  tetraquark states, see Table \ref{Assignment-Vector-P}.

In short, there are enough rooms to accommodate the existing $Y$ states above $4.2\,\rm{GeV}$.
From Table \ref{Assignment-Vector-P} and Table \ref{Interpretations-Table}, we can draw the conclusion tentatively that the $Y(4320/4360)$ and $Y(4390)$ maybe have more than one Fock components, which have an explicit P-wave between the diquark and antidiquark constituents or have an implicit P-wave in the diquark (or antidiquark) constituent; while the $Y(4660)$ maybe have  more than one Fock components, which have an implicit P-wave in the diquark (or antidiquark) constituent. Otherwise, there maybe exist several $Y$ states with almost degenerated masses but quite different quark structures.

In Table \ref{Compare-old}, we compare the present predictions with  the old calculations \cite{WangY4360Y4660-1803,Wang-tetra-formula,WangEPJC-1601-Mc,Vector-Tetra-WZG-4100}. From the Table, we can see explicitly that they are compatible with each other within uncertainties, or they have overlaps within uncertainties.  In this work, we  recalculate the contributions of the higher  dimensional vacuum condensates and acquire slightly  different analytical  expressions  compared with   the old calculations in Refs.\cite{WangY4360Y4660-1803,Wang-tetra-formula,WangEPJC-1601-Mc}, and correct some minor errors in the numerical calculations in Ref.\cite{Wang-tetra-formula}. Most importantly, we adopt the uniform/same constraint between the ground state masses $M_Y$ and continuum threshold parameters $s_0$, $\sqrt{s_0}=M_Y+0.5\pm 0.1\,\rm{GeV}$, for all the thirteen vector tetraquark states, and expect to acquire more robust/reliable QCD sum rules.

We can confront the vector hidden-charm tetraquark states predicted  in this work  to the experimental data in the future
 at the BESIII, LHCb, Belle II,  CEPC, FCC, ILC, which  maybe shed light on the nature of the exotic $X$, $Y$, $Z$ particles.
 We can investigate or search for the  neutral $Y_c$   tetraquark   states with the quantum numbers $J^{PC}=1^{--}$ and $1^{-+}$ through the two-body or three-body strong decays,
\begin{eqnarray}
  Y_c(1^{--}) &\to&  \chi_{c0}\rho/\omega \, ,\, J/\psi \pi^+\pi^- \, ,\,   J/\psi K\bar{K}\, ,\,  \eta_c\rho/\omega\, ,\, \chi_{c1}\rho/\omega\, , \nonumber\\
  Y_c(1^{-+}) &\to& J/\psi\rho/\omega \, ,\, h_c\rho/\omega \, .
\end{eqnarray}

\begin{sidewaystable}[thp]
\begin{center}
\begin{tabular}{|c|c|c|c|c|c|c|c|c|}\hline\hline
 $Y_c$                                                                     &$J^{PC}$  &$T^2(\rm{GeV}^2)$ &$\sqrt{s_0}(\rm GeV) $ &$\mu(\rm{GeV})$  &pole
&$M_Y (\rm{GeV})$   &$\lambda_Y (\rm{GeV}^5) $      \\ \hline

$[uc]_{P}[\overline{dc}]_{A}-[uc]_{A}[\overline{dc}]_{P}$                  &$1^{--}$  &$3.7-4.1$          &$5.15\pm0.10$          &$2.9$            &$(43-61)\%$ &$4.66\pm0.07$      &$(7.19\pm0.84)\times 10^{-2}$   \\

$[uc]_{P}[\overline{dc}]_{A}+[uc]_{A}[\overline{dc}]_{P}$                  &$1^{-+}$  &$3.7-4.1$          &$5.10\pm0.10$          &$2.8$            &$(42-60)\%$ &$4.61\pm0.07$      &$(6.69\pm0.80)\times 10^{-2}$   \\

$[uc]_{S}[\overline{dc}]_{V}+[uc]_{V}[\overline{dc}]_{S}$                  &$1^{--}$  &$3.2-3.6$          &$4.85\pm0.10$          &$2.4$            &$(42-62)\%$ &$4.35\pm0.08$      &$(4.32\pm0.61)\times 10^{-2}$   \\

$[uc]_{S}[\overline{dc}]_{V}-[uc]_{V}[\overline{dc}]_{S}$                  &$1^{-+}$  &$3.7-4.1$          &$5.15\pm0.10$          &$2.9$            &$(41-60)\%$ &$4.66\pm0.09$      &$(6.67\pm0.82)\times 10^{-2}$   \\

$[uc]_{\tilde{V}}[\overline{dc}]_{A}-[uc]_{A}[\overline{dc}]_{\tilde{V}}$  &$1^{--}$  &$3.6-4.0$          &$5.05\pm0.10$          &$2.7$            &$(42-60)\%$ &$4.53\pm0.07$      &$(1.03\pm0.14)\times 10^{-1}$  \\

$[uc]_{\tilde{V}}[\overline{dc}]_{A}+[uc]_{A}[\overline{dc}]_{\tilde{V}}$  &$1^{-+}$  &$3.7-4.1$          &$5.15\pm0.10$          &$2.9$            &$(41-60)\%$ &$4.65\pm0.08$      &$(1.13\pm0.15)\times 10^{-1}$  \\

$[uc]_{\tilde{A}}[\overline{dc}]_{V}+[uc]_{V}[\overline{dc}]_{\tilde{A}}$  &$1^{--}$  &$3.5-3.9$          &$5.00\pm0.10$          &$2.6$            &$(42-61)\%$  &$4.48\pm0.08$      &$(9.47\pm1.27)\times 10^{-2}$   \\

$[uc]_{\tilde{A}}[\overline{dc}]_{V}-[uc]_{V}[\overline{dc}]_{\tilde{A}}$  &$1^{-+}$  &$3.6-4.0$          &$5.05\pm0.10$          &$2.7$            &$(42-61)\%$ &$4.55\pm0.07$      &$(1.06\pm0.14)\times 10^{-1}$  \\

$[uc]_{S}[\overline{dc}]_{\tilde{V}}-[uc]_{\tilde{V}}[\overline{dc}]_{S}$  &$1^{--}$  &$3.4-3.8$          &$5.00\pm0.10$          &$2.6$            &$(42-61)\%$ &$4.50\pm0.09$      &$(4.78\pm0.66)\times 10^{-2}$   \\

$[uc]_{S}[\overline{dc}]_{\tilde{V}}+[uc]_{\tilde{V}}[\overline{dc}]_{S}$  &$1^{-+}$  &$3.4-3.8$          &$5.00\pm0.10$          &$2.6$            &$(42-61)\%$ &$4.50\pm0.09$      &$(4.79\pm0.66)\times 10^{-2}$  \\

$[uc]_{P}[\overline{dc}]_{\tilde{A}}-[uc]_{\tilde{A}}[\overline{dc}]_{P}$  &$1^{--}$  &$3.7-4.1$          &$5.10\pm0.10$          &$2.8$            &$(43-61)\%$  &$4.60\pm0.07$      &$(6.32\pm0.74)\times 10^{-2}$  \\

$[uc]_{P}[\overline{dc}]_{\tilde{A}}+[uc]_{\tilde{A}}[\overline{dc}]_{P}$  &$1^{-+}$  &$3.7-4.1$          &$5.10\pm0.10$          &$2.8$            &$(43-61)\%$  &$4.61\pm0.08$      &$(6.36\pm0.74)\times 10^{-2}$  \\

$[uc]_{A}[\overline{dc}]_{A}$                                              &$1^{--}$  &$3.8-4.2$          &$5.20\pm0.10$          &$3.0$            &$(42-60)\%$ &$4.69\pm0.08$      &$(6.65\pm0.81)\times 10^{-2}$  \\

\hline\hline
\end{tabular}
\end{center}
\caption{ The Borel parameters, continuum threshold parameters, energy scales of the QCD spectral densities,  pole contributions,  masses and pole residues  for the ground state vector hidden-charm tetraquark states, where the constraint in Eq.\eqref{formula-Regge} is satisfied. }\label{BorelP}
\end{sidewaystable}

\begin{sidewaystable}[thp]
\begin{center}
\begin{tabular}{|c|c|c|c|c|c|c|c|c|}\hline\hline
 $Y_c$                                                                     &$J^{PC}$  &$T^2(\rm{GeV}^2)$  &$\sqrt{s_0}(\rm GeV) $  &pole         &$\widehat{M}_Y (\rm{GeV})$   &$\widehat{\lambda}_Y (\rm{GeV}^5) $         \\ \hline

$[uc]_{P}[\overline{dc}]_{A}-[uc]_{A}[\overline{dc}]_{P}$                  &$1^{--}$  &$4.0-4.6$          &$5.70\pm0.10$           &$(40-62)\%$  &$5.20\pm0.07$      &$(7.95\pm 1.05)\times 10^{-2}$       \\

$[uc]_{P}[\overline{dc}]_{A}+[uc]_{A}[\overline{dc}]_{P}$                  &$1^{-+}$  &$4.0-4.6$          &$5.65\pm0.10$           &$(40-61)\%$  &$5.14\pm0.08$      &$(7.39\pm 1.01)\times 10^{-2}$ \\

$[uc]_{S}[\overline{dc}]_{V}+[uc]_{V}[\overline{dc}]_{S}$                  &$1^{--}$  &$3.3-3.7$          &$5.25\pm0.10$           &$(41-61)\%$  &$4.75\pm0.09$      &$(4.02\pm 0.69)\times 10^{-2}$   \\

$[uc]_{S}[\overline{dc}]_{V}-[uc]_{V}[\overline{dc}]_{S}$                  &$1^{-+}$  &$3.6-4.1$          &$5.50\pm0.10$           &$(40-62)\%$  &$5.00\pm0.10$      &$(5.49\pm 0.87)\times 10^{-2}$  \\

$[uc]_{\tilde{V}}[\overline{dc}]_{A}-[uc]_{A}[\overline{dc}]_{\tilde{V}}$  &$1^{--}$  &$3.7-4.2$          &$5.50\pm0.10$           &$(40-61)\%$  &$4.99\pm0.09$      &$(1.01\pm0.16)\times 10^{-1}$  \\

$[uc]_{\tilde{V}}[\overline{dc}]_{A}+[uc]_{A}[\overline{dc}]_{\tilde{V}}$  &$1^{-+}$  &$3.6-4.1$          &$5.50\pm0.10$           &$(40-62)\%$  &$4.99\pm0.09$      &$(0.93\pm0.16)\times 10^{-1}$  \\

$[uc]_{\tilde{A}}[\overline{dc}]_{V}+[uc]_{V}[\overline{dc}]_{\tilde{A}}$  &$1^{--}$  &$3.5-4.0$          &$5.40\pm0.10$           &$(40-62)\%$  &$4.90\pm0.08$      &$(8.78\pm 1.42)\times 10^{-2}$  \\

$[uc]_{\tilde{A}}[\overline{dc}]_{V}-[uc]_{V}[\overline{dc}]_{\tilde{A}}$  &$1^{-+}$  &$3.8-4.3$          &$5.55\pm0.10$           &$(41-61)\%$  &$5.05\pm0.08$      &$(1.11\pm 0.16)\times 10^{-1}$  \\

$[uc]_{S}[\overline{dc}]_{\tilde{V}}-[uc]_{\tilde{V}}[\overline{dc}]_{S}$  &$1^{--}$  &$3.2-3.6$          &$5.25\pm0.10$           &$(41-62)\%$  &$4.75\pm0.11$      &$(3.44\pm0.66)\times 10^{-2}$  \\

$[uc]_{S}[\overline{dc}]_{\tilde{V}}+[uc]_{\tilde{V}}[\overline{dc}]_{S}$  &$1^{-+}$  &$3.2-3.6$          &$5.25\pm0.10$           &$(41-62)\%$  &$4.75\pm0.11$      &$(3.45\pm0.66)\times 10^{-2}$  \\

$[uc]_{P}[\overline{dc}]_{\tilde{A}}-[uc]_{\tilde{A}}[\overline{dc}]_{P}$  &$1^{--}$  &$4.0-4.6$          &$5.65\pm0.10$           &$(40-61)\%$  &$5.15\pm0.07$      &$(7.10\pm0.94)\times 10^{-2}$  \\

$[uc]_{P}[\overline{dc}]_{\tilde{A}}+[uc]_{\tilde{A}}[\overline{dc}]_{P}$  &$1^{-+}$  &$4.0-4.6$          &$5.65\pm0.10$           &$(40-61)\%$  &$5.16\pm0.07$      &$(7.14\pm0.93)\times 10^{-2}$   \\

$[uc]_{A}[\overline{dc}]_{A}$                                              &$1^{--}$  &$3.7-4.2$          &$5.55\pm0.10$           &$(41-62)\%$  &$5.05\pm0.09$      &$(5.58\pm 0.87)\times 10^{-2}$  \\

\hline\hline
\end{tabular}
\end{center}
\caption{ The Borel parameters, continuum threshold parameters, pole contributions, masses and pole residues for the
ground state vector hidden-charm tetraquark states with the energy scales $\mu=1\,\rm{GeV}$,  where the constraint in Eq.\eqref{formula-Regge} is not satisfied. }\label{BorelP-1GeV}
\end{sidewaystable}

\begin{table}
\begin{center}
\begin{tabular}{|c|c|c|c|c|c|c|c|c|}\hline\hline
  $Y_c$                                                                    & $J^{PC}$  & $M_Y (\rm{GeV})$  & Interpretations         &$\widehat{M}_Y (\rm{GeV})$ \\ \hline

$[uc]_{P}[\overline{dc}]_{A}-[uc]_{A}[\overline{dc}]_{P}$                  &$1^{--}$   &$4.66\pm0.07$      & ?\,\,$Y(4660)$      &$5.20\pm0.07$       \\

$[uc]_{P}[\overline{dc}]_{A}+[uc]_{A}[\overline{dc}]_{P}$                  &$1^{-+}$   &$4.61\pm0.07$      &                     &$5.14\pm0.08$     \\

$[uc]_{S}[\overline{dc}]_{V}+[uc]_{V}[\overline{dc}]_{S}$                  &$1^{--}$   &$4.35\pm0.08$      & ?\,\,$Y(4360/4390)$ &$4.75\pm0.09$     \\

$[uc]_{S}[\overline{dc}]_{V}-[uc]_{V}[\overline{dc}]_{S}$                  &$1^{-+}$   &$4.66\pm0.09$      &                     &$5.00\pm0.10$    \\

$[uc]_{\tilde{V}}[\overline{dc}]_{A}-[uc]_{A}[\overline{dc}]_{\tilde{V}}$  &$1^{--}$   &$4.53\pm0.07$      &                     &$4.99\pm0.09$     \\

$[uc]_{\tilde{V}}[\overline{dc}]_{A}+[uc]_{A}[\overline{dc}]_{\tilde{V}}$  &$1^{-+}$   &$4.65\pm0.08$      &                     &$4.99\pm0.09$      \\

$[uc]_{\tilde{A}}[\overline{dc}]_{V}+[uc]_{V}[\overline{dc}]_{\tilde{A}}$  &$1^{--}$   &$4.48\pm0.08$      &                     &$4.90\pm0.08$       \\

$[uc]_{\tilde{A}}[\overline{dc}]_{V}-[uc]_{V}[\overline{dc}]_{\tilde{A}}$  &$1^{-+}$   &$4.55\pm0.07$      &                     &$5.05\pm0.08$      \\

$[uc]_{S}[\overline{dc}]_{\tilde{V}}-[uc]_{\tilde{V}}[\overline{dc}]_{S}$  &$1^{--}$   &$4.50\pm0.09$      &                     &$4.75\pm0.11$     \\

$[uc]_{S}[\overline{dc}]_{\tilde{V}}+[uc]_{\tilde{V}}[\overline{dc}]_{S}$  &$1^{-+}$   &$4.50\pm0.09$      &                     &$4.75\pm0.11$      \\

$[uc]_{P}[\overline{dc}]_{\tilde{A}}-[uc]_{\tilde{A}}[\overline{dc}]_{P}$  &$1^{--}$   &$4.60\pm0.07$      &                     &$5.15\pm0.07$     \\

$[uc]_{P}[\overline{dc}]_{\tilde{A}}+[uc]_{\tilde{A}}[\overline{dc}]_{P}$  &$1^{-+}$   &$4.61\pm0.08$      &                     &$5.16\pm0.07$     \\

$[uc]_{A}[\overline{dc}]_{A}$                                              &$1^{--}$   &$4.69\pm0.08$      & ?\,\,$Y(4660)$      &$5.05\pm0.09$     \\
\hline\hline
\end{tabular}
\end{center}
\caption{ The possible interpretations of the ground state vector hidden-charm tetraquark states, the isospin limit is implied, the
vector tetraquark states $\bar{c}c\bar{d}u$, $\bar{c}c\bar{u}d$, $\bar{c}c\frac{\bar{u}u-\bar{d}d}{\sqrt{2}}$ and
$\bar{c}c\frac{\bar{u}u+\bar{d}d}{\sqrt{2}}$ have almost degenerated masses. }\label{Interpretations-Table}
\end{table}

\begin{table}
\begin{center}
\begin{tabular}{|c|c|c|c|c|c|c|c|c|}\hline\hline
  $Y_c$                                                                    & $J^{PC}$  &$M_Y (\rm{GeV})$   &$M_Y (\rm{GeV})$ (Old Work)         \\ \hline

$[uc]_{P}[\overline{dc}]_{A}-[uc]_{A}[\overline{dc}]_{P}$                  &$1^{--}$   &$4.66\pm0.07$      &$4.59 \pm 0.08$ \cite{WangY4360Y4660-1803}; $4.66^{+0.17}_{-0.10}$ \cite{Wang-tetra-formula}    \\

$[uc]_{P}[\overline{dc}]_{A}+[uc]_{A}[\overline{dc}]_{P}$                  &$1^{-+}$   &$4.61\pm0.07$      &$4.57^{+0.12}_{-0.08}$   \cite{Wang-tetra-formula}    \\

$[uc]_{S}[\overline{dc}]_{V}+[uc]_{V}[\overline{dc}]_{S}$                  &$1^{--}$   &$4.35\pm0.08$      &$4.34\pm 0.08$ \cite{WangY4360Y4660-1803} \\

$[uc]_{S}[\overline{dc}]_{V}-[uc]_{V}[\overline{dc}]_{S}$                  &$1^{-+}$   &$4.66\pm0.09$      &          \\

$[uc]_{\tilde{V}}[\overline{dc}]_{A}-[uc]_{A}[\overline{dc}]_{\tilde{V}}$  &$1^{--}$   &$4.53\pm0.07$      &          \\

$[uc]_{\tilde{V}}[\overline{dc}]_{A}+[uc]_{A}[\overline{dc}]_{\tilde{V}}$  &$1^{-+}$   &$4.65\pm0.08$      &        \\

$[uc]_{\tilde{A}}[\overline{dc}]_{V}+[uc]_{V}[\overline{dc}]_{\tilde{A}}$  &$1^{--}$   &$4.48\pm0.08$      &        \\

$[uc]_{\tilde{A}}[\overline{dc}]_{V}-[uc]_{V}[\overline{dc}]_{\tilde{A}}$  &$1^{-+}$   &$4.55\pm0.07$      &       \\

$[uc]_{S}[\overline{dc}]_{\tilde{V}}-[uc]_{\tilde{V}}[\overline{dc}]_{S}$  &$1^{--}$   &$4.50\pm0.09$      &$4.61\pm0.08$  \cite{Vector-Tetra-WZG-4100}     \\

$[uc]_{S}[\overline{dc}]_{\tilde{V}}+[uc]_{\tilde{V}}[\overline{dc}]_{S}$  &$1^{-+}$   &$4.50\pm0.09$      &            \\

$[uc]_{P}[\overline{dc}]_{\tilde{A}}-[uc]_{\tilde{A}}[\overline{dc}]_{P}$  &$1^{--}$   &$4.60\pm0.07$      &            \\

$[uc]_{P}[\overline{dc}]_{\tilde{A}}+[uc]_{\tilde{A}}[\overline{dc}]_{P}$  &$1^{-+}$   &$4.61\pm0.08$      &              \\

$[uc]_{A}[\overline{dc}]_{A}$                                              &$1^{--}$   &$4.69\pm0.08$      &$4.66\pm0.09$ \cite{WangEPJC-1601-Mc}      \\
\hline\hline
\end{tabular}
\end{center}
\caption{ The masses of the  ground state vector hidden-charm tetraquark states compared with the old calculations. }\label{Compare-old}
\end{table}

\section{Conclusion}
In the present work,  we adopt the scalar, pseudoscalar,  vector,  axialvector and tensor   (anti)diquark operators  as the elementary  building blocks    to construct vector and tensor  four-quark currents as many as possible without introducing an explicit P-wave (in other words, the negative-parity of the (anti)diquarks embodies the P-wave effect in an implicit way),  and explore the mass spectrum of the vector hidden-charm tetraquark states via the QCD sum rules  comprehensively,  and revisit the interpretations of the existing  $Y$ states in the  scenario of vector tetraquark  states. We acquire eight original QCD sum rules and update the calculations of the five old QCD sum rules. In addition, we resort to  the energy scale formula to enhance the pole contributions and improve the convergent behaviors of the operator product expansion, and we should bear in mind that the predictions are rather sensitive to the particular  energy scales which obey the uniform/same  constraint.
The present predictions support identifying the $Y(4360/4390)$ as the $[uc]_{S}[\overline{dc}]_{V}+[uc]_{V}[\overline{dc}]_{S}$ hidden-charm tetraquark state with the  quantum numbers $J^{PC}=1^{--}$ and identifying the
$Y(4660)$ as the $[uc]_{P}[\overline{dc}]_{A}-[uc]_{A}[\overline{dc}]_{P}$  or  $[uc]_{A}[\overline{dc}]_{A}$ hidden-charm tetraquark state with the quantum numbers  $J^{PC}=1^{--}$.

 We take account of our previous works on the vector tetraquark states with an explicit P-wave between the diquark and antidiquark constituents, and acquire the conclusion tentatively that the $Y(4320/4360)$,  $Y(4390)$ and $Y(4660)$ maybe have more than one Fock components. The $Y(4320/4360)$ and $Y(4390)$ maybe have an explicit P-wave between the diquark and antidiquark constituents or an implicit P-wave in the diquark (or antidiquark) constituent;  the $Y(4660)$ maybe have an implicit P-wave in the diquark (or antidiquark) constituent. Otherwise  there maybe exist more than one $Y$ states with almost degenerated masses but quite different quark configurations.

All in all, we can  accommodate all the exotic $Y$ states above $4.2\,\rm{GeV}$ in a consistent way.
We can confront the predicted vector hidden-charm tetraquark states   to the experimental data  at the BESIII, LHCb, Belle II,  CEPC, FCC, ILC in the future.

\section*{Acknowledgements}
This  work is supported by National Natural Science Foundation, Grant Number  11775079.

\end{document}